# Sporting the government: Twitter as a window into sportspersons' engagement with causes in India and USA

Dibyendu Mishra, Microsoft Research India

Ronojoy Sen, National University of Singapore

Joyojeet Pal, Microsoft Research India

## ABSTRACT

With the ubiquitous reach of social media, influencers are increasingly central to articulation of political agendas on a range of topics. We curate a sample of tweets from the 200 most followed sportspersons in India and the United States respectively since 2019, map their connections with politicians, and visualize their engagements with key topics online. We find significant differences between the ways in which Indian and US sportspersons engage with politics online – while leading Indian sportspersons tend to align closely with the ruling party and engage minimally in dissent, American sportspersons engage with a range of political issues and are willing to publicly criticize politicians or policy. Our findings suggest that the ownership and governmental control of sports impact public stances on issues that professional sportspersons are willing to engage in online. It might also be inferred, depending upon the government of the day, that the costs of speaking up against the state and the government in power have different socio-economic costs in the US and India.

## 1. INTRODUCTION

The conversations on Information and Communications Technology for Development (ICTD) in the last decade have increasingly engaged beyond the traditional domains such as healthcare, education, agriculture etc. to issues around gender rights, democratization, and media. Perhaps the most significant change in the technology landscape in the Global South in recent years has been the meteoric rise in access to mobile devices, which in turn has led to a corresponding increase in social media access. This has made social media an important artifact of direct news and media access among citizens in many parts of the world who were formerly restricted in their access to broadcast and interpersonal communications, and arguably, in turn engaging with the democratic process.

But has an increase in access to social media created new forms of information regulation? In this paper, we study a foundational question in social media and societal change – whether the development of, and access to technological tools changes the quality of democratic discourse, specifically through the behavior of online influencers. At the heart of this question is the extent to which citizens in a nation-state can consume the output from free and responsible media, but also exercise their own right to free speech, particularly when that speech is oppositional to the government in power. To interrogate this, we designed a two-country study, comparing leading sportspersons in India and the United States. In India where most sportspersons are part of a system of national team representation or sporting associations where the state plays a significant part, the relationship of sportspersons with the state is relatively close whereas in the United States, most public figures in sporting are part of private enterprise in which the state exercises minimal agency. What a public figure feels free to discuss in public fora, like Twitter and Facebook, can be seen as reflective of how they feel, as influential citizens, about engaging in public discourse. These questions therefore are in turn related to the broader issues around social media and the consolidation of political power in nation states.

## 2. RELATED WORK

### 2.1 Sportspersons and Spokespersons

It is often said that sports and politics should ideally not mix. This, however, has never been true of sport at the highest levels, especially when sportspersons are representing their nation. George Orwell had perhaps overstated this idea when he famously described sport, in the context of Cold War politics, as "war minus the shooting." (Orwell, 1945) A more nuanced analysis on the connection between sport and nation has been offered by the great historian E.J. Hobsbawm: "What has made sport uniquely effective as a medium for inculcating national feelings, at all events for males, is the ease with which even the least political or public individual can identify with the nation as symbolized by young persons excelling at something practically every man wants to be good at. The imagined community of millions seems more real as a team of eleven named people." (Kesavan, 2021) The Olympics are perhaps the best example of an international sporting contest, which was conceived by its founder Pierre de Coubertin as a means to bring about fraternal relations between nations, but has become "synonymous with nationalism" (Kesavan, 2021). Hence, the heartburn in India over its perennial under-performance in the Olympics and the exultation after the 2021 Tokyo Olympics when India did better than usual.

Sports are part of the national imaginary and an aspirational frame, in part because it is among a small number of activities, despite its gendered inequities, that engage citizens across class. Consequently, citizens often do not grudge obscenely high remuneration for professional sportspersons (Seippel, et al., 2018), and hero worship of leading sportspersons is common across countries (Berg, 1998). Sport has traditionally enjoyed an important space in both identity and politics (Bowman, 2015) and often serves as common language for integration in diverse societies (Zec & Paunović, 2015). Nation states have sporting events as large spectacles, hold up individual sportspersons as international heroes, and have indeed even gone to war over sports (Chirinos, 2018).

Sporting brands have been highly monetized in both professional club sports and international competitive sports such as the Olympic games (Smart, 2018). There is an entire economic ecosystem around star sportspersons including sponsorships, tourism (Gammon, 2014) , and entire industries dedicated to the management and articulation of sportspersons' brands (Zhou, et al., 2020), which impact an individual or team's behavior and its perceived value (Yoon & Shin, 2017). Sportspersons' brands have consequences for the social alliances and causes they undertake (Madrigal, 2000), and are increasingly an important part of corporate social responsibility (Smith & Westerbeek, 2007).

Countries also hold close sports that are considered central to their national identity among large parts of their population. Thus, team sports like Rugby in South Africa or New Zealand (Black & Nauright, 1998), Soccer in much of Latin America or Europe (Seippel, 2017) (Bar-On, 1997), Ice Hockey in Canada or Norway (Watson, 2017), or Cricket in South Asia (Perera, 2000) are central to cultural practices and public discourse. Individual sportspersons, including those outside of some of these sports who capture public attention, are often massively important public figures who are influential in a range of spheres outside of their immediate expertise. Depending on the country and sport, individual sportspersons' public lives may extend well beyond their immediate playing years (James & Nadan, 2020). Sportspersons' influence in politics is enabled by the existence of a political system that allows for it. Having popular positions can further a sportsperson's immediate career but impact their post-sporting career through alignment with adversarial framings (Boykoff & Carrington, 2020).

The United States offers an important set of cases around the intersection of sports and politics. A critical part of religious identity in the United States, sports are also places where faith identities may be publicly performed by major sporting stars (Parker & Watson, 2015). In the US, political engagement by sportspersons has been particularly central to racial politics, marked by key moments of protest such as boxer Muhammad Ali's refusal to fight in Vietnam (Gorsevski & Butterworth, 2011) or the raised fist protest at the 1968 Mexico City Olympic games by sprinters Tommie Smith and John Carlos, which while iconic, eventually led to the two athletes being widely criticized and ostracized by the sporting community (Hartmann, 2003). While some positions – like that of sportspersons expressing strong popular positions like the praying, patriotic American football quarterback Tim Tebow can be projected as heroic (Butterworth, 2013) (Watts, 2014), other positions antithetical to the dominant institutional or political discourse have always been fraught. Nowhere was this more pronounced than in the systematic economic attack on football quarterback Colin Kaepernick, who unlike the praying Tebow, took the knee while the American national anthem was playing to protest racism, a perceived slight that lost him his professional career (Boykoff & Carrington, 2020) (Chaplin & de Oca, 2019). While sportspersons who follow the lead of pioneering protesters do not always face vociferous or systematic boycott, they nonetheless deal with distractions by way of public rebuke, trolling, losses of economic engagements, and threats to bodily harm (Schmidt, et al., 2019).

In India, it is well known that sporting contests, especially in cricket, between India and Pakistan bring forth intense nationalist feelings (Appadurai, 1995; Guha, 2003; Sen, 2015). Conversely, cricket is also seen as a tool to foster better relations between India and Pakistan, which has led to coining of the term “cricket diplomacy.” On both counts, a heavy burden is placed on the cricketers. The India-Pakistan sporting contests are also a test of loyalty, particularly for India’s large Muslim minority population.

## 2.2 Sportspersons on Social Media

Social media has become very important for sportspersons to express their multi-faceted selves beyond sports and interact with their fans. A study that coded tweets based on their motives from 101 athletes in USA from various sports found that about 62% of them were either used to interact directly with fans or for sharing non-sports related content (Hambrick, et al., 2010). A similar study (Pegoraro, 2010) on top American athletes over a 7-day period from various sports confirmed that sportspersons used Twitter significantly to interact with fans or share personal and non-sports content. Another study (Sanderson, 2013) also explored how early career athletes in major leagues in USA also posted stories and pictures from their personal life and made pop culture references along with their training regimens to build their identities in the public sphere.

Athletes are also able to benefit as businesspersons through their activity on social media. A study of Twitter behavior showed that by commenting on important events, sportspersons are able to increase their following and extend their brand value (Korzynski & Paniagua, 2016) They also point out that sportspersons with less success on the field are also able to build a better brand for themselves using these strategies. Some studies have also focused specifically on the engagement sportspersons have with sociopolitical issues. Shmargad (Shmargad, 2018) found that low-ranking candidates and challengers in the 2016 US elections were more likely to gain high voter percentages if they were retweeted by highly influential users like celebrity sportspersons. Another study (Yan, et al., 2018) utilized network analysis around social issue hashtags to also find that athletes display organizational dynamics on Twitter, while other studies have investigated visibility and gendered economies of sport to find differences in the content of tweets of top athletes (Lebel & Danylchuk, 2012) (Toffoletti & Thorpe, 2018). While there is a body of work that has investigated the impact of a newly emerging group of influencers i.e. online influencers on social media platforms and the content they disseminate (Bakshy, et al., 2011) (Lalani, et al., 2019) (Dubois & Gaffney, 2014) (Zarei, et al., 2020) (Cha, et al., 2010), there are no large-scale comparative analyses of influential athletes' tweets, specifically analyzing their politics.

## 3. DATA

In the absence of an off the shelf list of most highly followed athletes in USA and India, we used a ‘wisdom of experts’ approach to curate a list on Twitter (Ghosh, et al., 2013). While there have been other efforts which utilize friendship graphs, LDA clusters based on profile information and a combination of both (Weng, et al., 2010) (Pal & Counts, 2011), finding a highly precise list of top followed accounts on a specific topic remains a challenge. One study (Ghosh, et al., 2012) showed that by utilizing the ‘lists’ feature on Twitter, which allows users to group other users based on what they think their topic of expertise is, we can build such a list with higher precision. However, the method also relies on crawling the data of a large number of Twitter users similar to the methods outlined earlier. They also found high subjectivity in these curations on certain topics. To address the scalability issue and reduce noise due to subjectivity, we manually curate lists maintained by topic experts like sports journalists and news media. We crowdsourced 33 such lists for Indian sportspersons and 47 lists for American sportspersons. We then combined the members of all lists, sorted them by their followers count and annotated these users to find the top 200 athletes based on followers count from both USA and India.

A part of our study also explores the direct engagements of sportspersons with influential politicians in their respective countries. To this end, we curate accounts of 5,000 highly influential accounts of politicians in India and USA from pre-existing databases (Panda, et al., 2020) (Anonymous, 2021). We first include legislators and senators from the US and elected representatives, specifically members of Parliament and members of state legislative assemblies from India and further append the list with highly followed politicians’ accounts. For our study we use Twitter's Academic API to collect tweets from January, 2019 to April, 2021 for all sportspersons’ accounts. We retrieve 102,878 tweets from Indian sportspersons and 158,705 tweets from American sportspersons in this time frame. Our logic for using this time frame was that it included a five-month buffer before and after general elections in both countries, which helps craft the activity on influencer engagement during, before, and after a key political event.

## 4. ANALYSIS AND FINDINGS

We present our findings in two subsections. The first one deals with direct engagements with the top influential politicians in their country and the second deals with engagement with important political events and topics in general.

### 4.1 Interactions with Politicians

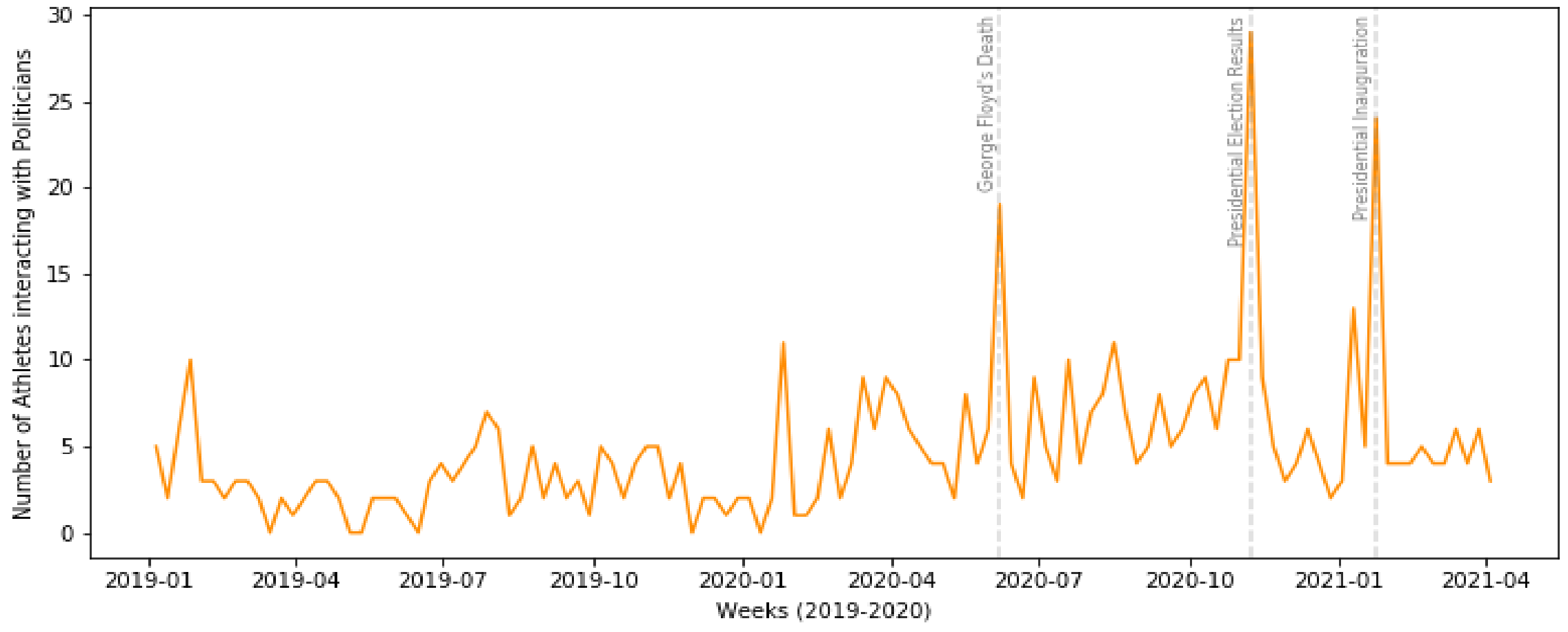

**Fig 1 (a):** A weekly timeline of US Athletes engagements with US Politicians

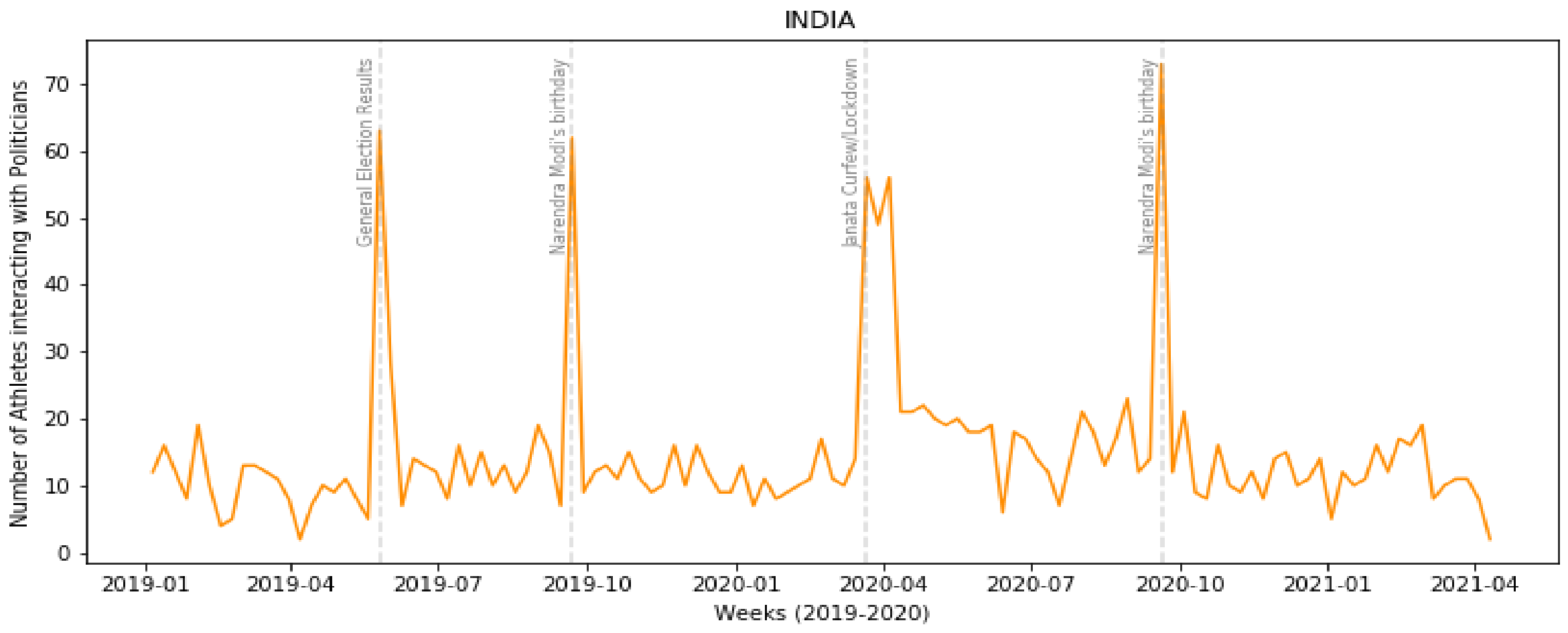

**Fig 1 (b):** A weekly timeline of Indian Athletes engagements with Indian Politicians

In Figures 1 (a) and 1 (b) we chart the total number of tweets by sportspersons directly interacting with politicians on Twitter by way of mentions, quotes and replies. In general, we see a high number of tweets interacting with politicians on the Indian side in comparison to those on the American side. However, the higher representation on the Indian size does not indicate greater political commentary. We see that the major spikes on the American side are around George Floyd's death, the BLM (Black Lives Matter) protests and the US Presidential elections. On the other hand, on the Indian side, we see major spikes around the initial COVID-19 lockdown, the 2019 General Election results and oddly on Prime Minister Narendra Modi's birthday in both 2019 and 2020. Over a third of all the top 200 Indian sportspersons engaged with Modi on his birthday in 2020.

To analyze the contents of the tweets that engage with politicians we plot the chatter plots (figure 2(a) and 2(b)) of the most frequent words occurring in these conversations. We see that a good number of tweets engage in thanking the politicians in both countries, indicating that most number conversations are largely courtesy conversations. However, as we analyze other words and the tweets related to them we see more American sportspersons calling people to 'vote', engaging with the 'black' lives matter movement, talking about the 'rights' of people and raising issues around the 'economy'. It is also important that while the American sportspersons were engaging with the BLM movement, the government in place in the United States was generally seen as opposed to the movement. In contrast, there is very little engagement of Indian sportspersons with positions that are explicitly opposed to the ruling party.

Indian sportspersons' most frequent engagements are wishes to politicians on their birthdays. Stylistically, the semantic construction of most of these tweets is formal, using terms like 'sir' and 'hounorable'. Another stylistic pattern we note is that the tweets often do not directly converse with politicians but have them mentioned at the end of tweets, as if to alert a politician that a certain tweet was posted. Another pattern distinct in India is the use of hashtags that are being popularized by the government – a common example of which is 'indiafightscorona', which was largely used by members of the ruling party. We also see a mobilization around other state-sponsored social media campaigns including 'fitindiamovement'.

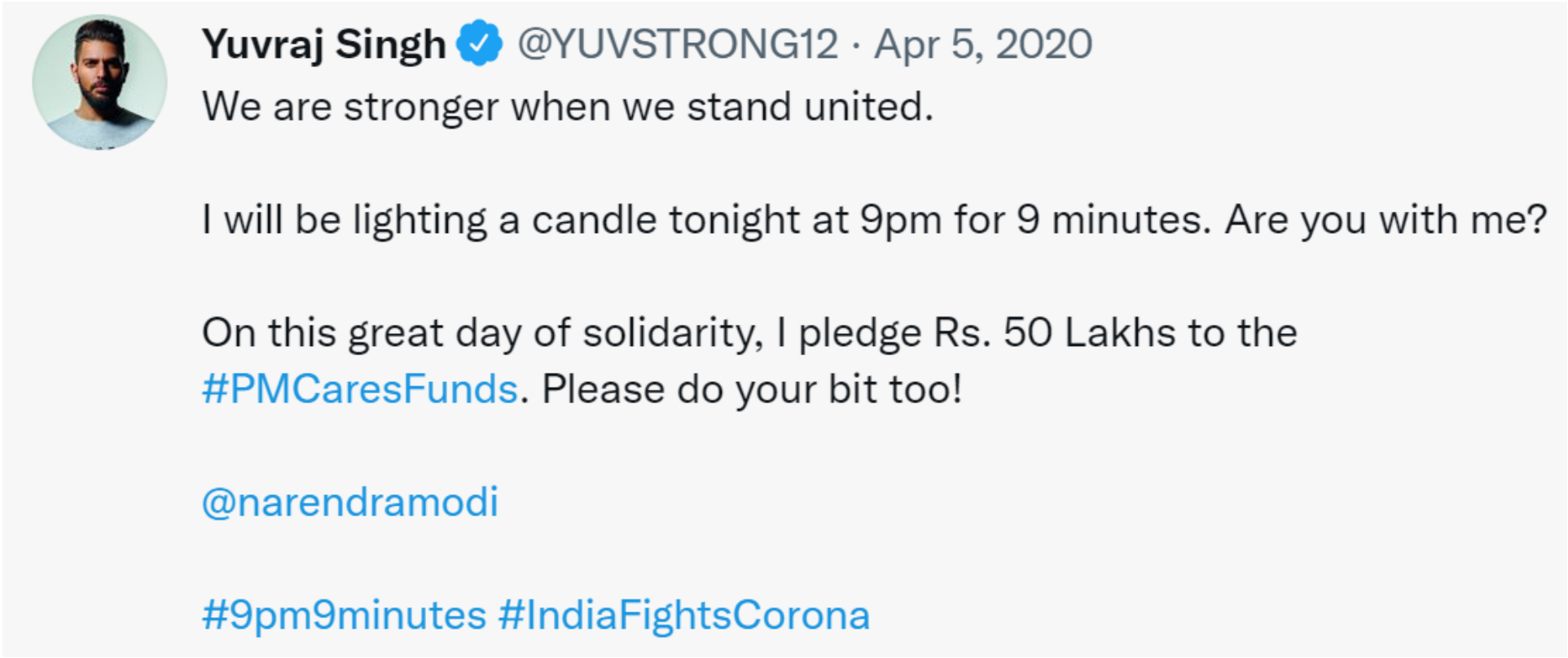

**Figure 2**: Cricketer Yuvraj Singh calls for participation in the #9pm9minutes and pledges his donation to the PMCaresFund while mentioning Narendra Modi at the end of the tweet

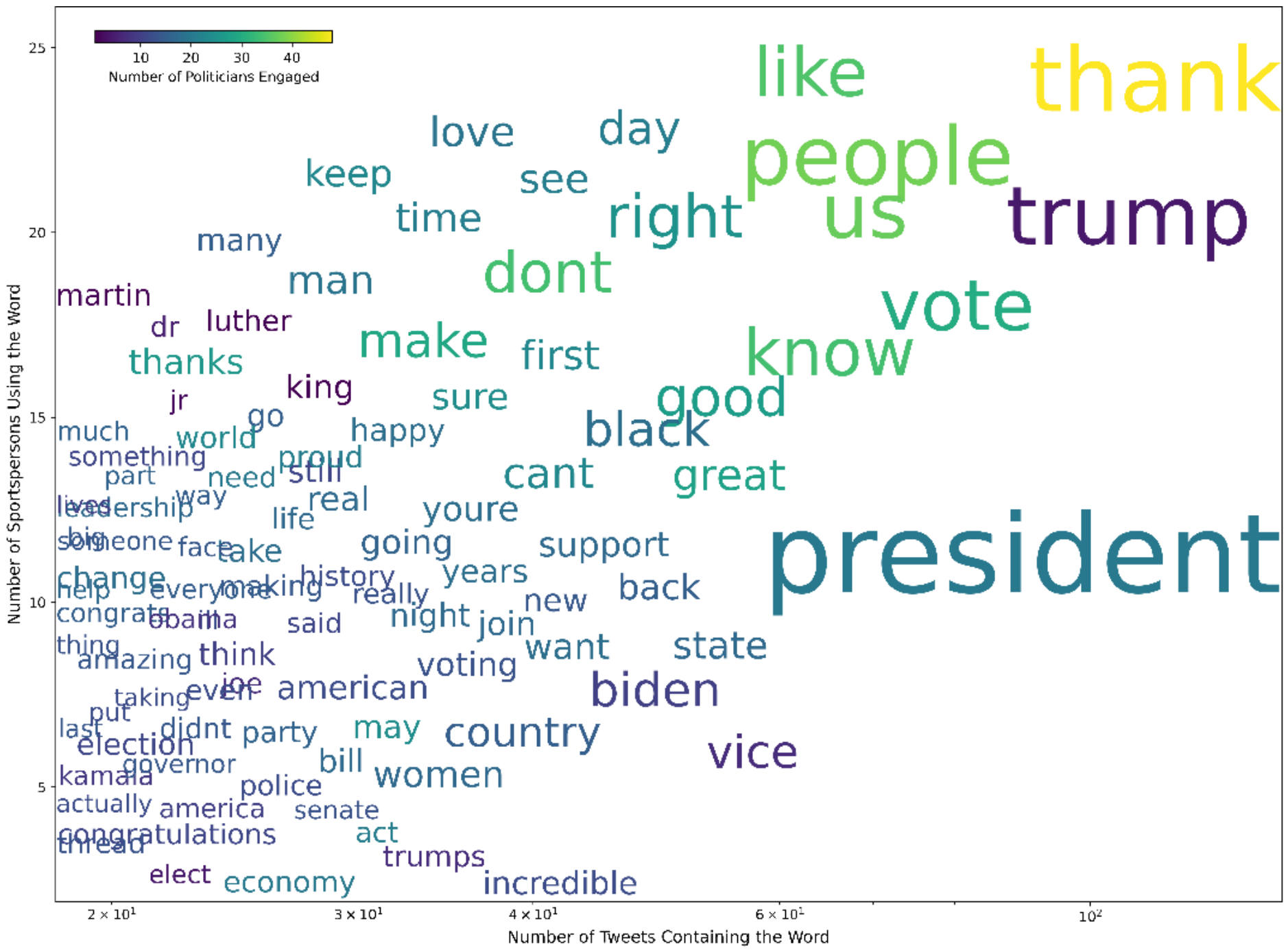

**Fig 3(a)**: Chatter plot of Engagements of American Sportspersons with Politicians

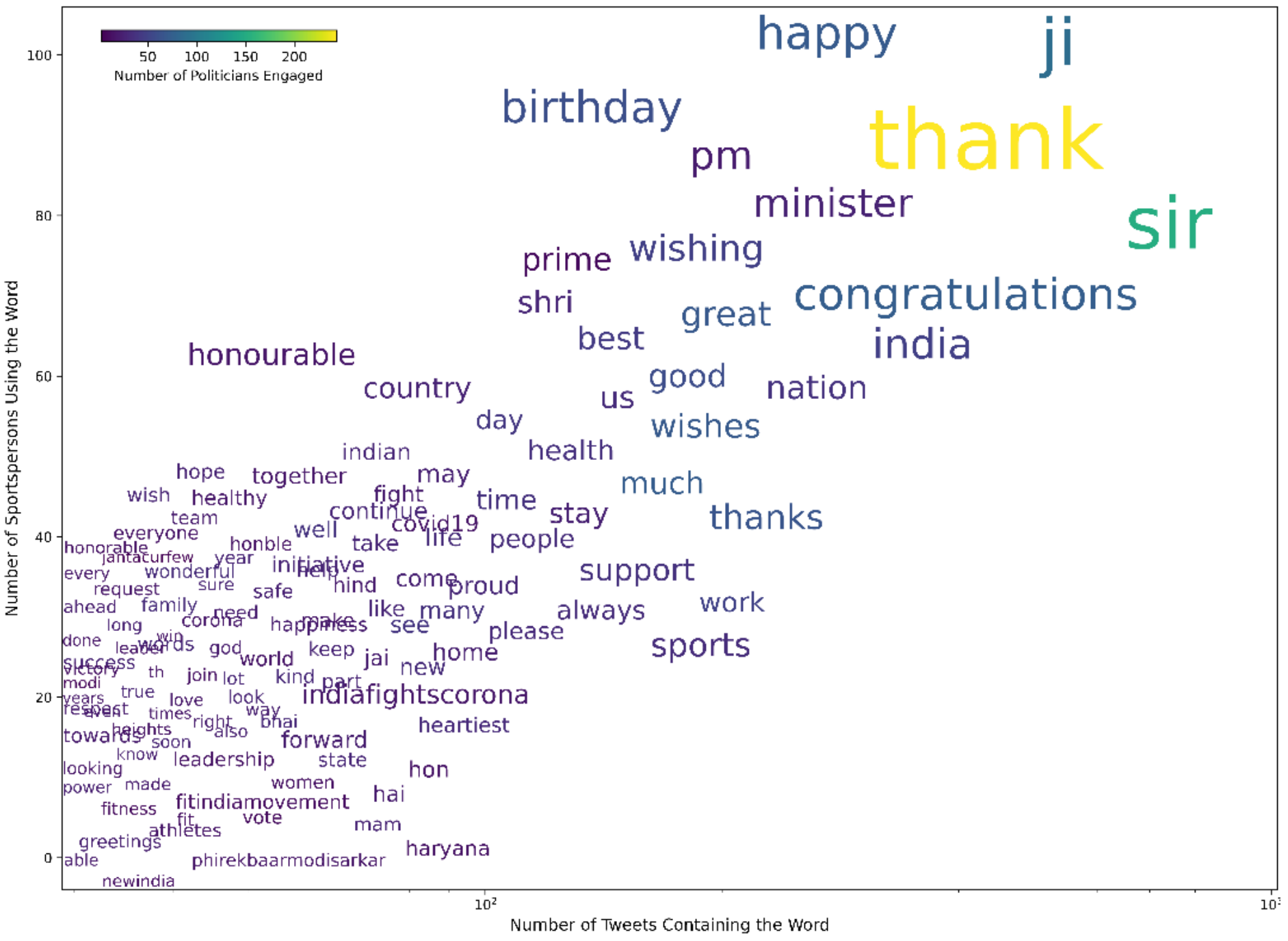

**Fig 3(b)**: Chatter plot of Engagements of Indian Sportspersons with Politicians

## 4.2 Political Tweeting

In this section we analyze tweets that do not necessarily engage directly with politicians, yet talk about issues of socio-political importance. We do this by selecting topics around issues and events that took centerstage in either country in the temporal window that we are studying. On the Indian side, we look at protest events centered around the Citizenship Amendment Act (CAA) and Indian Farm Bills passed in the Parliament in September 2020. Along with these we also look at conversations of Indian athletes on prominent topics around caste, gender and sexual orientation. In the end we also discuss COVID-19 and 2019 General Election related awareness campaigns and tweets. Similarly on the US side, we discuss prominent protests against racial discrimination like Black Lives Matter and #StopAAPIHate. We then move on to discussing movements and issues around gender and sexual orientation and awareness campaigns and tweets around COVID-19 and the 2020 Presidential elections.

Towards this, we first define a list of seed terms for each event/topic. We then train a Word2Vec (Mikolov, et al., 2013) model on the tweets containing these terms to get a continuous distributed vector representation of terms and capture the context surrounding each term in the tweets. Using a cosine similarity based shortlisting procedure (Vijayaraghavan, et al., 2021), we add terms most similar to the seed terms and that are most representative of the issue or topic. We capture the final list of tweets related to each topic or event by simply filtering them using this query set. A selection from the expanded query sets for each topic/event is presented below:

| **Topics** | **Keywords and Hashtags (stemmed)** |
|---|---|
| Protests around race and ethnicity | #blacklivesmatter, #georgefloyd, #stopasianhate |
| On issues related to gender and sexual orientation | lgbt, #metoo, #timesup, #equalpay, #wagegap, rape, sexist,#orangetheworld, #lovewins |
| On elections and voting | election, #morethanavote, #registeredandready, vote |
| On COVID-19 | covid, mask, vaccin, socaildistanc, corona, pandemic |

**Table 1(a)**: A selection of Keywords and Hashtags for each event/issue in USA

| **Topics** | **Keywords and Hashtags (stemmed)** |
|---|---|
| Prominent Protest Events | #farmersprotest, #caa, #nrc, #indiatogehter |
| On issues of caste and ethnicity | reservation, caste, bahujan, dalit, hathras, #rajputboy |
| On issues related to gender and sexual orientation | rape, hergametoo, #womenpower, #metoo, hathras, #pridemonth, #bharatkilaxmi |
| On elections and voting | #votekar, #everyvotecounts, #deshkamahatyohar, elections |
| On COVID-19 | covid, pandemic, vaccin, #janatacurfew, corona |

**Table 1(b):** A selection of Keywords and Hashtags for each event/issue in India

### 4.2.1 American Sportspersons

In the US, we find that NBA players are particularly vocal on social media including Lebron James, Jamal Crawford, and Kyle Kuzuma who tweet consistently and vociferously on issues of race and polarization in the United States. Two antidiscrimination-related issues, *Black Lives Matter* and *Stop Asian Hate* feature significantly in the messaging from sportspersons.

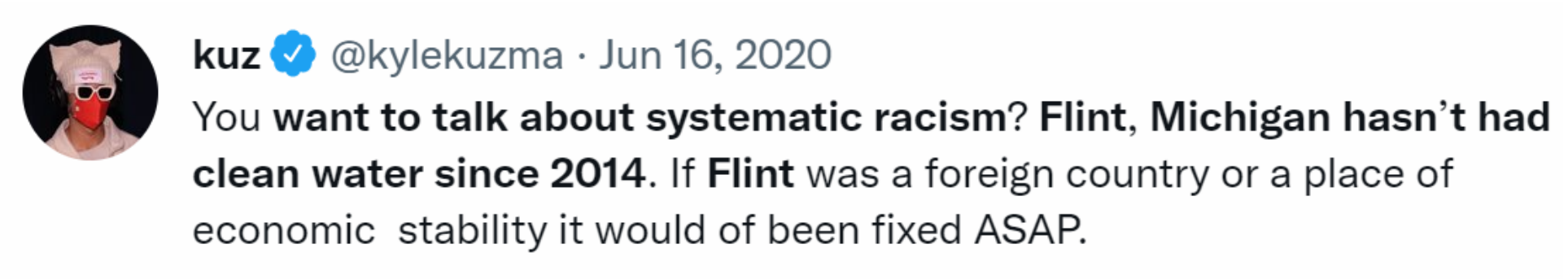

**Figure 4**: NBA player Kyle Kuzma on systemic racism in America

While events that grabbed public imagination and went viral on social media were key drivers of sportsperson engagements, conversations on broader social inequities on gender, race, and orientation found purchase among a range of sportspersons across gender, race, and sport. A number of female athletes spoke up on pay disparities between male and female sportspersons. The 2019 presidential election in the US was also a major point of convergence with a vast majority of posts from athletes being opposed to the incumbent president Donald Trump. US sportspersons move away from mere tokenism – there are frontal attacks on systemic racism, calls to action aimed at voters, and attacks on specific institutions seen as complicit in discrimination including the Presidency and the police.

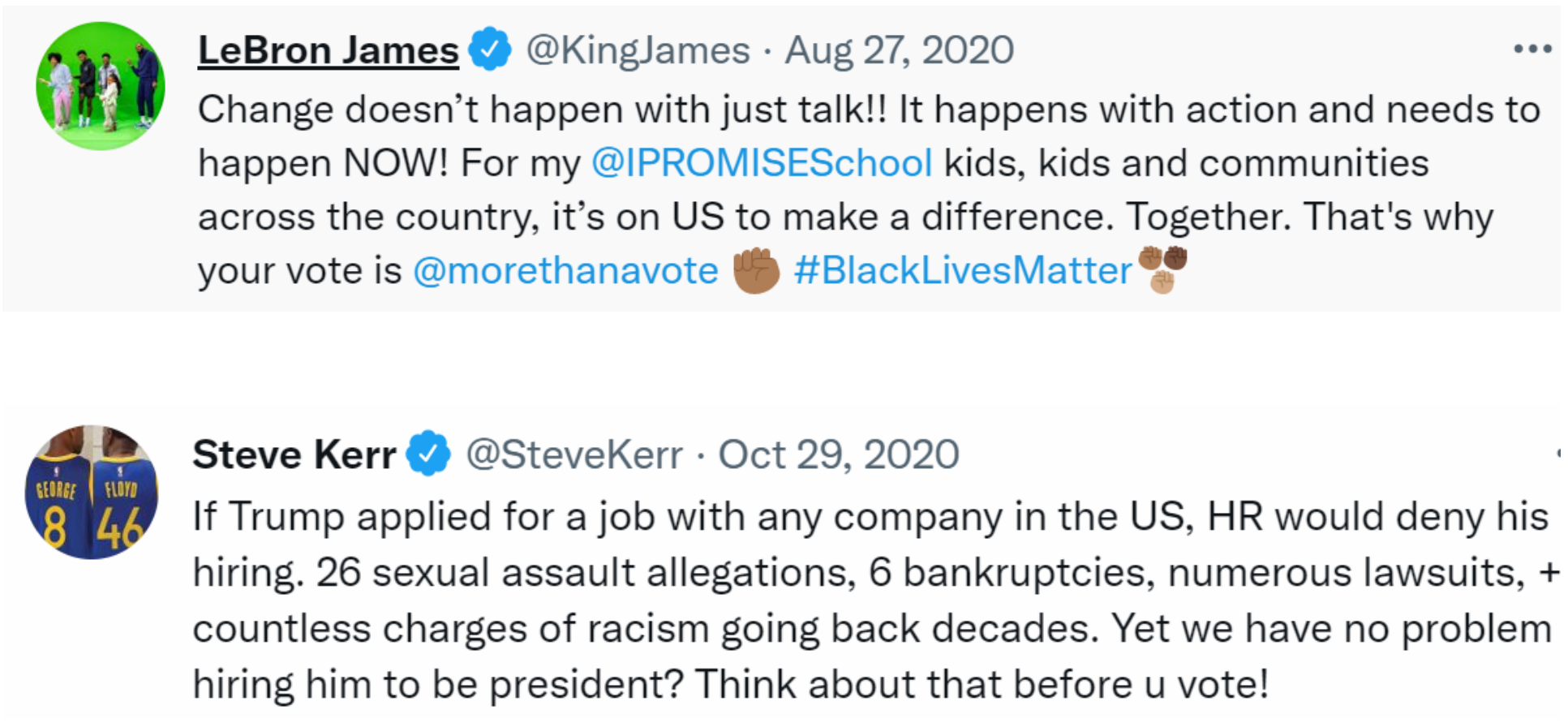

**Figure 5:** Direct calls to Action by Citizens from NBA player LeBron James and coach Seve Kerr

Pro basketball player and YouTube influencer Jeremy Lin who messaged prominently on issues of Asian Hate, was joined by Tennis Grand Slam winner Naomi Osaka in messaging about racially motivated attacks in the aftermath of COVID. A differentiating factor of these tweets is that they do not simply condemn a certain kind of agreeably despicable behavior, such as attacks on a set of people, but rather that they specifically call out blame to the institutions that enable such hate or the role of individual leaders or segments of society in these acts.

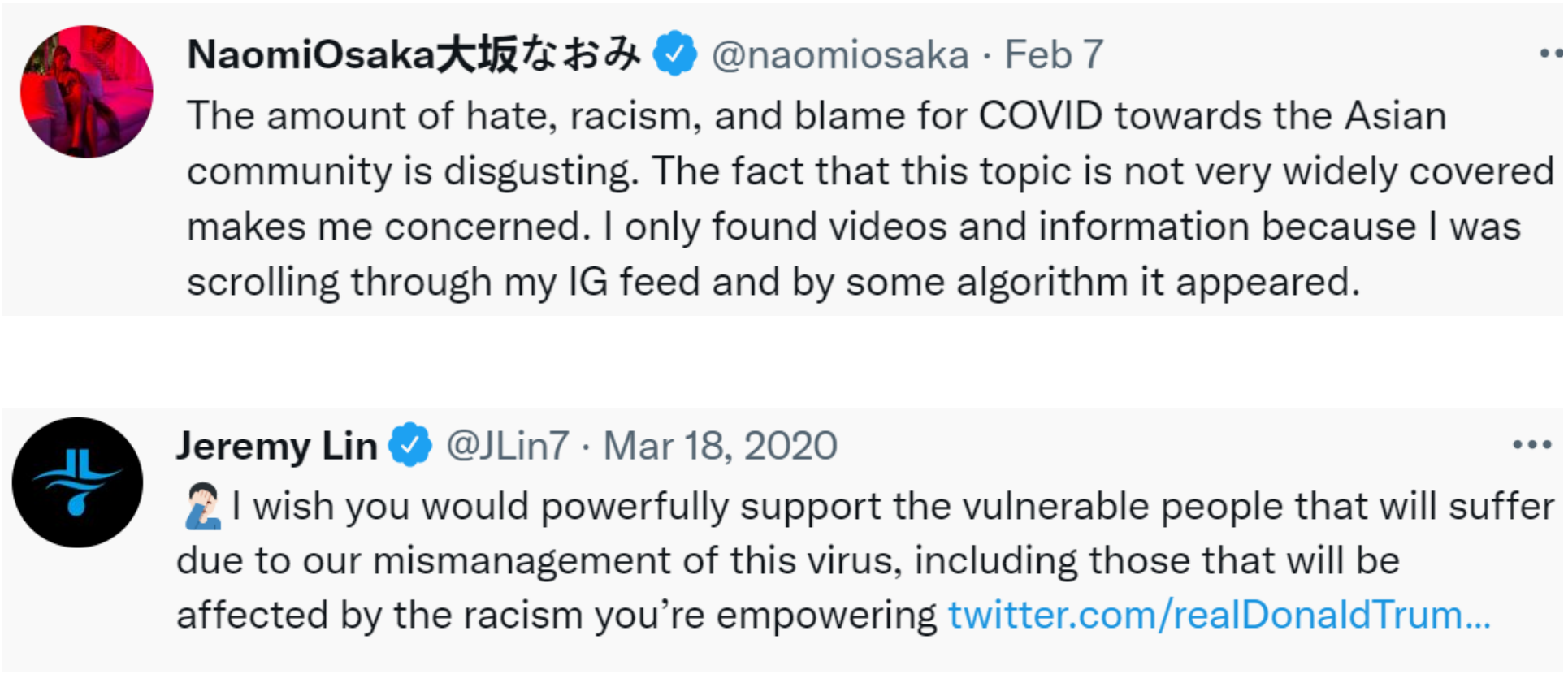

**Figure 6:** Tweets on Asian American hate incidents related to COVID

American sportspersons also engaged heavily on topics around masking up and social distancing. This was despite a significant portion of the population, generally intersecting with conservative voters, which declined to wear masks and being out of work with a large share of sporting activities being shut down.

### 4.2.2 Indian Sportspersons

As with US sportspersons, in India too, we see sportspersons urging citizens to vote through hashtag campaigns like #votekar, #festivalofdemocracy and #getinked. Unlike in the US, where sportspersons explicitly noted who they felt their audiences should vote for, Indian sportspersons were less direct. However, a number of the calls for people to vote were direct responses to Prime Minister Narendra Modi's personalized calls to action mentioning these sportspersons, which by extension implicitly supported the incumbent party although a small number of sportspersons like shot putter Deepa Malik, Badminton player Saina Nehwal gave direct endorsements to Modi by using his campaign hashtags. The rest of the tweets congratulated Modi and his party for their eventual victory.

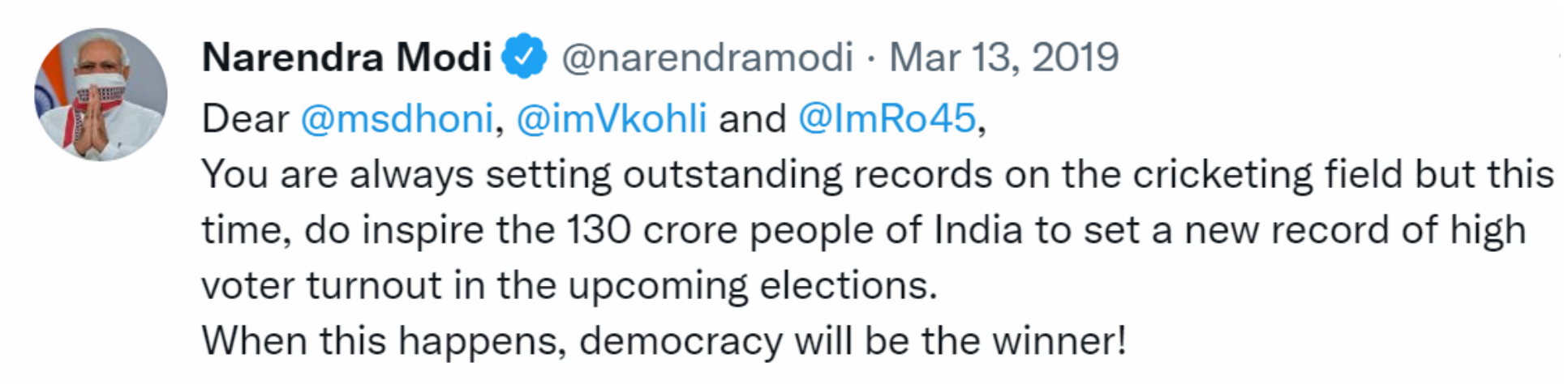

**Figure 7:** PM Modi directly calling upon influential cricketers to inspire citizens to vote

The first wave of COVID-19 saw sportspersons asking citizens to obey the nationwide lockdown and stay home by engaging with hashtags put forth by the government including #jantacurfew and #indiafightscorona. A specific case in which we see a flurry of activity is the trending of the ruling party's initiative, #9baje9minute, which urged citizens to light candles in their homes for nine minutes on April 6, 2020. Sportspersons were also important voices in pledging and asking donations for the #PMCaresFund, a fund for the incumbent Prime Minister's COVID relief efforts. Unlike in the US, where sportspersons tweeted about pandemic mismanagement, we see little if any evidence of that in India, despite the massive toll that the crisis took in India.

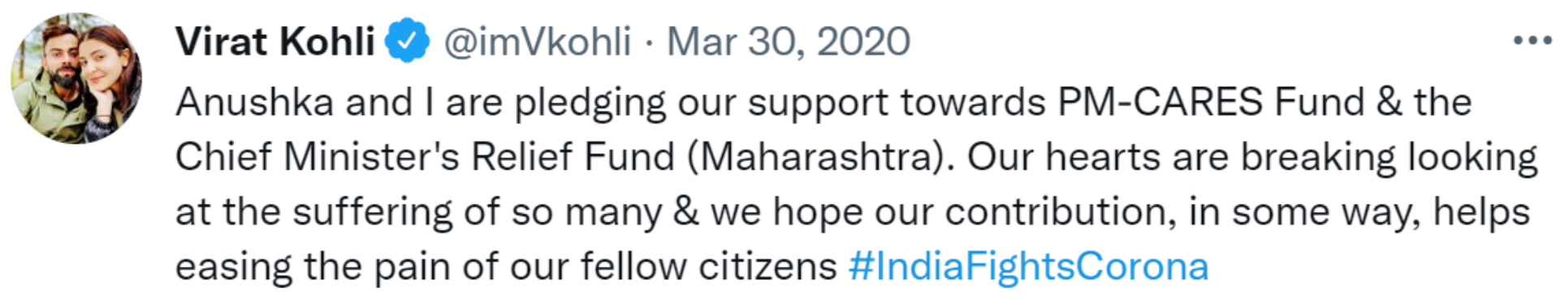

**Figure 8:** Indian cricket captain Virat Kohli pledging support to the PM-CARES Fund

One issue that saw some confrontational messaging by sportspersons was the controversial Citizenship Amendment Act (CAA) and National Register of Citizens (NRC) which sought to reframe citizenship laws as per religious affiliation. The protests against CAA-NRC did get tweeted about: while boxer Manoj Kumar tweeted against the protests calling them anti-Hindu, conspiratorial and declared his complete support for the government's move, a handful, including former cricketers Irfan Pathan, Sanjay Manjrekar and Aakash Chopra, Badminton player Jwala Gutta, and Olympian Shiva Keshavan condemned the violence against protesting students by the police. However, no sportsperson barring former footballer CK Vineeth came out in support of the protests.

On discrimination in society, there was very little active engagement by sportspersons. While there is research that shows that discriminated groups such as Bahujans are underrepresented and discriminated against in sports (Barua, 2019; Shantha, 2017; Sarmah, 2020), there was no nationwide anti-caste movement on social media comparable to what we see with BLM in the US. When incidents of caste violence took place, such as the rape of a Dalit woman that sparked protests across the country, sportspersons including influential cricketers like Virat Kohli and Rohit Sharma did take to their handles to condemn the violence but did not address it as a caste-based issue. On the issue of affirmative action, sportspersons like basketball player Divya Singh and boxer Manoj Kumar spoke out against caste-based reservations, calling for its removal, calling it a 'murder of merit' and caste-based atrocities as a thing of the past. At least two sportspersons from privileged castes like cricketers Ravindra Jadeja and Shikhar Dhawan actively celebrated their caste heritage. Despite their silence on discrimination in India, a small number of sportspersons including cricketer Hardik Pandya and swimmer Sumit Nagal took to their handles to support the #BlackLivesMatter movement.

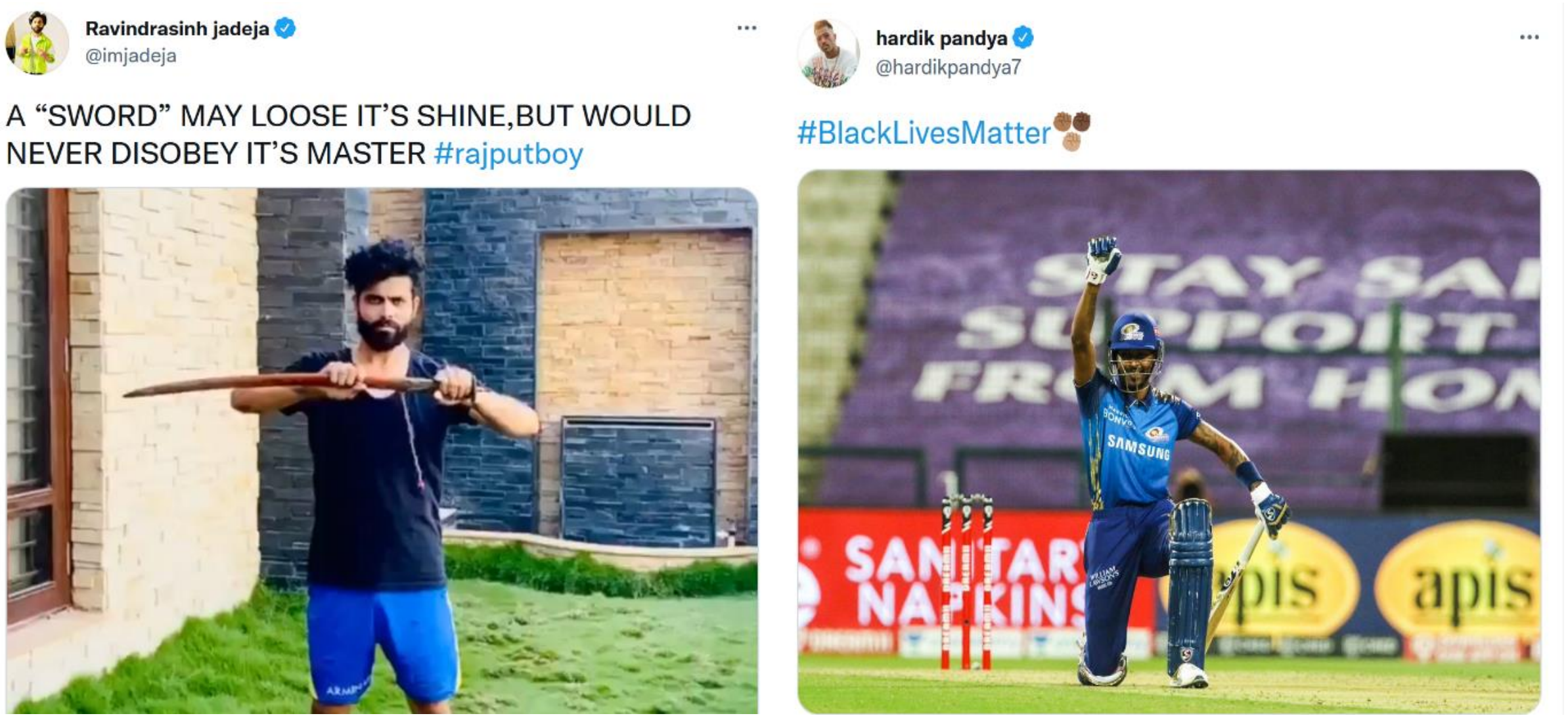

**Figure 9**: (Left)Cricketer Ravindra Jadeja wielding a sword to show his caste pride (Right) Cricketer Hardik Pandya Kneels down in solidarity with the BLM movement

On gender issues, a large number of tweets participated in government sanctioned hashtag campaigns like #bharatkilaxmi, though on at least one issue that saw action with US sportspersons – that of the gender pay gap - cricketer Mithali Raj, Olympian Dipa Karmakar, and cricketer Aakash Chopra spoke up. Sportspersons remained silent on LGBTQIA+ rights with a lone track and field athlete Dutee Chand actively talking about it on her feed and celebrating pride month.

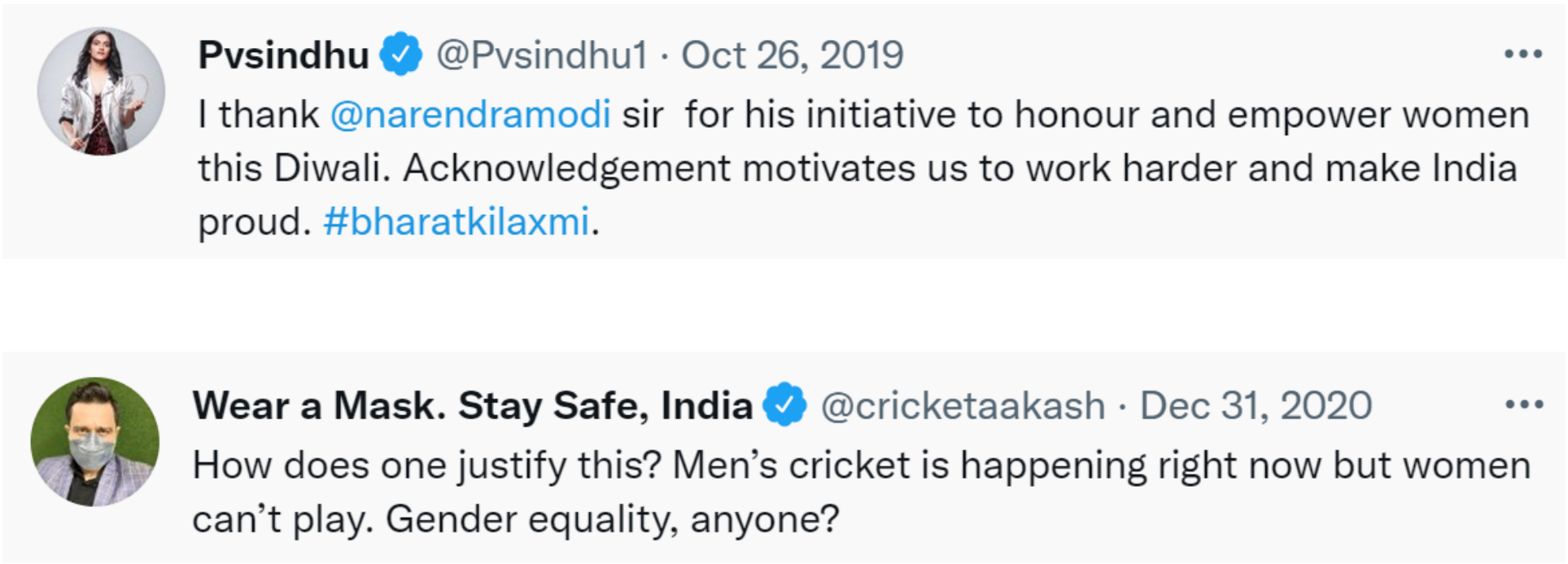

**Figure 10**: (Top) Badminton player PV Sindhu on the #bharatkilaxmi campaign (Bottom) Aakash Chopra calling attention to unequal opportunities for women cricketers

Perhaps the most important political hot topic on social media in India was the 2020 Farmers' Protest. Unlike other events, the side opposed to the incumbent government in this case was the farmers, who are both politically powerful, and culturally revered. Furthermore, the sporting establishment draws a large share of its stars from the two states of Punjab and Haryana, which were key in protesting the farm laws, and prominent sportspersons from these states including wrestlers Bajrang Punia, Sakshi Malik and Vinesh Phogat, and hockey players like Poonam Rani Malik and Sandeep Singh all spoke up against the central government for police action, and supported the farmer's demands. The more influential accounts, a majority of whom are cricketers, remained silent on the issue barring a few like the former cricketers Harbhajan Singh and Wasim Jaffer. However, after prominent singer and businessperson Rihanna tweeted calling attention to the issue, Indian twitter saw a flurry of tweets almost in a coordinated fashion (Mishra, et al., 2021) from its most influential sportstars like cricketers Virat Kohli and Sachin Tendulkar and badminton stars like Saina Nehwal which called for India to be together and for the non-interference of 'external forcers', referring to Rihanna, Meena Harris, Mia Khalifa, and Greta Thunberg, among others, and trended the hashtag '#IndiaAgainstPropaganda'. A few others like basketball player Divya Singh questioned the integrity of the protests and called for arrests of the farmers.

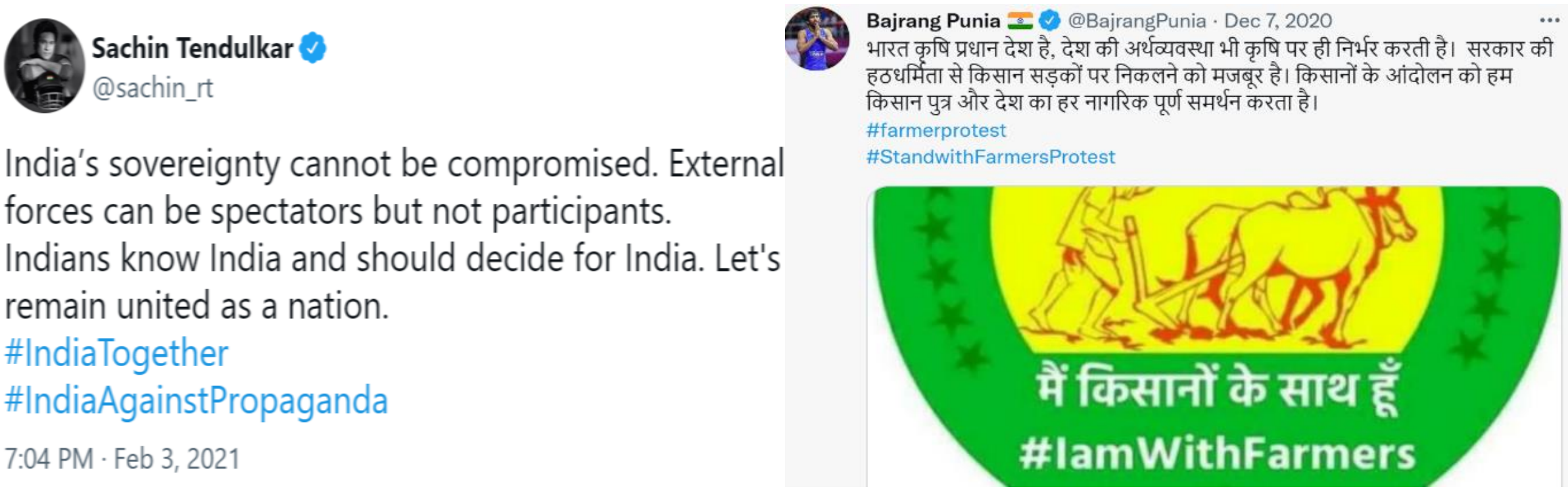

**Figure 11:** (Left) Star Cricketer Sachin Tendulkar condemning Rihanna's tweet on the farmers' protest (Right) Wrestler Bajrang Punia condemning the government in support of the farmers' protest

## CONCLUSION

Unlike in the US, where most political messaging was antagonistic towards Trump, rather than celebratory of Biden, there is practically no frontal attack on Modi by any of the Indian sportspersons. On the contrary, the massive engagement of sportspersons on Modi's birthday has the effect of 'paying tribute.' We see that sportspersons are among the most influential amplifiers of campaign hashtags, systematically employed at times when the government has initiatives it wishes to get people enthused about. This suggests an effective capture of sportspersons' social media accounts in the interests of state initiatives and governance. US athletes on the other hand tend to engage more directly with social and political issues that are presumably their own or through their independent endorsement relationships, since we see no comparable pattern of structured engagement in state-led campaigns. While we do see the less influential sportspersons, who are mostly non-cricketers and more dependent on the state in terms of funding and opportunities such as representing the country, actively engaging in issues like the farmers' protest, the silence of star cricketers, who hold private club contracts, offers an important case in point. Only a handful of retired cricketers engage in any messaging that is explicitly opposed to the government's positions. It is presumable that many Indian sportspersons do believe in the precise causes that the government puts out and hence their engagements cannot be called outright sycophancy.  Nonetheless, even for sports that do not have explicit state intervention, sportspersons fail to leverage their huge social media influence to raise issues of the marginalized. In short, the differences we notice between the public commentary by sports influencers in the two systems serve as a window into the ways political culture can influence, guide, or command sporting culture and vice versa.